\documentclass[11pt,preprint]{aastex}
\usepackage{graphicx}
\usepackage{amssymb}
\usepackage{amsmath}
\usepackage{hyperref}

\def\gtrsim{\mathrel{\hbox{\rlap{\hbox{\lower4pt\hbox{$\sim$}}}\hbox{$>$}}}}
\def\lesssim{\mathrel{\hbox{\rlap{\hbox{\lower4pt\hbox{$\sim$}}}\hbox{$<$}}}}
\def\gtrsim{\mathrel{\hbox{\rlap{\hbox{\lower4pt\hbox{$\sim$}}}\hbox{$>$}}}}

\def\farcm{\hbox{$.\!\!^{\prime}$}}

\begin{document}
\title{Multi--wavelength study of HESS J1741$-$302}

\author{Jeremy Hare\altaffilmark{1}, Blagoy Rangelov\altaffilmark{1},  Eda Sonbas\altaffilmark{1,2},  Oleg Kargaltsev\altaffilmark{1}, and Igor Volkov\altaffilmark{1,3}}
\altaffiltext{1}{Department of Physics, The George Washington University, 725 21st St, NW, Washington, DC 20052}
\altaffiltext{2}{Department of Physics, University of Adiyaman, 02040 Adiyaman, Turkey}
\altaffiltext{3}{University of Maryland, College Park, MD 20742 }
\email{jeh86@gwu.edu}

\slugcomment{The Astrophysical Journal, in press}
\shorttitle{HESS~J1741$-$302}

\begin{abstract}

We present the results of two {\sl Chandra X-ray Observatory} ({\sl CXO}) observations of TeV $\gamma$-ray source HESS J1741--302. We investigate whether there is any connection between HESS~J1741$-$302 and the sources seen at lower energies. One of the brightest X-ray sources in the HESS J1741--302 field, CXOU~J174112.1$-$302908, appears to be associated with a low-mass star (possibly representing  a quiescent  LMXB or CV), hence, it is unlikely to be a source of TeV $\gamma$-rays. In the same field we have potentially detected X-rays from WR 98a, which is likely to be a colliding wind binary with  massive stars. No TeV emission has been reported so far from such systems although predictions have been made. Finally, we found that the previously reported {\sl Suzaku} source, \emph{Suzaku} J1740.5--3014 (which is not covered by the {\sl CXO} observations), appears to be a hard X-ray source detected by {\sl INTERGAL}  ISGRI, which supports the magnetized CV classification but makes its association with the TeV emission unlikely. The young pulsar PSR B1737-30, so far  undetected in X-rays and projected on the sky near the CV, may be the contributor of relativistic particles responsible for the TeV emission.

\end{abstract}


\section{Introduction}

The High Energy Stereoscopic System (H.E.S.S.) has revealed many TeV $\gamma$-ray sources in the Galactic plane \citep{2005Sci...307.1938A}. Roughly half of the total number of sources in the Galactic plane ($\sim$90) have been firmly associated with high mass X-ray binaries (HMXBs), shell-type supernova remnants (SNRs), and pulsar-wind nebulae (PWNe). The latter appear to dominate the overall population of Galactic TeV sources   (see \citealt{2013arXiv1305.2552K} for review). 
There is a substantial number of unidentified very high energy (VHE; detected above 1 TeV) sources ($\sim$20), some of which have plausible multi-wavelength (MW) counterparts (e.g., nearby young pulsars) but 
the associations have not been confidently established yet (e.g., there is no X-ray PWN, X-ray and TeV emission do not correlate, offsets from pulsars are too large). Finally, there are  few  
($\sim$7) VHE sources that belong to so-called ``dark accelerators" because they have no plausible counterparts at any other wavelength (see \citealt{2013arXiv1305.2552K}).  It was suggested that these VHE sources could be relic PWNe of older undetected pulsars (see e.g.,  \citealt{2009arXiv0906.2644D}, \citealt{2011arXiv1111.1634T}, \citealt{2013ApJ...773..139V}).

HESS J1741-302, located near the Galactic center, was discovered during a  H.E.S.S. survey of the Galactic plane \citep{2008AIPC.1085..249T}. This source was detected with a significance of 8.1$\sigma$ in 143.5 hours of observations \citep{2009arXiv0907.0574T}.  The TeV image of the source appears to exhibit two hot spots which, due to poor statistics, could not be definitively claimed to be two independent sources \citep{2009arXiv0907.0574T}. In this paper we tentatively label these hot spots as HESS J1741-302A and HESS J1741-302B (hereafter J1741A and J1741B, respectively; see Figure 1). We define the HESS J1741A region as a $4\farcm2$ circle at a position of R.A.=17$^{h}$41$^{m}$40$^{s}$ and Decl. = $-$30$^{\circ}$05$''$00$'$ and the HESS J1741B region as a $6\farcm4$ circle at a position of R.A.=17$^{h}$41$^{m}$17$^{s}$ and Decl.=$-$30$^{\circ}$23$''$00$'$, based on Figure 5 from \cite{2009arXiv0907.0574T}. Two scenarios have been put forth to describe the emission from HESS J1741-302, namely hadronic gamma-ray production via interaction of cosmic rays with molecular clouds in the region, or a PWN associated with the offset, yet relatively young and powerful, pulsar B1737--30 \citep{2009arXiv0907.0574T}.

\begin{figure*}
\begin{center}
\includegraphics[scale=0.29, angle=270]{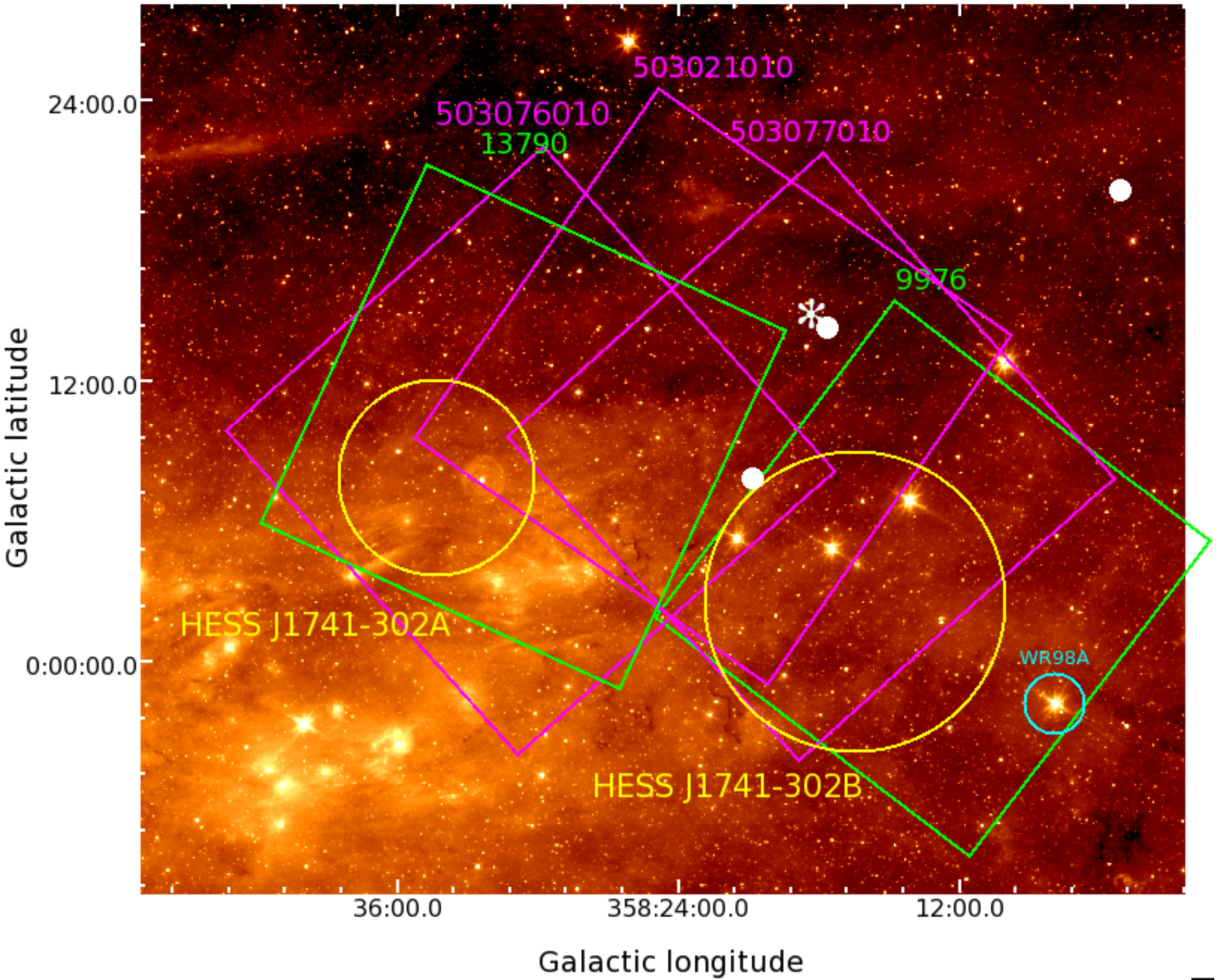}
\includegraphics[scale=0.29, angle=270]{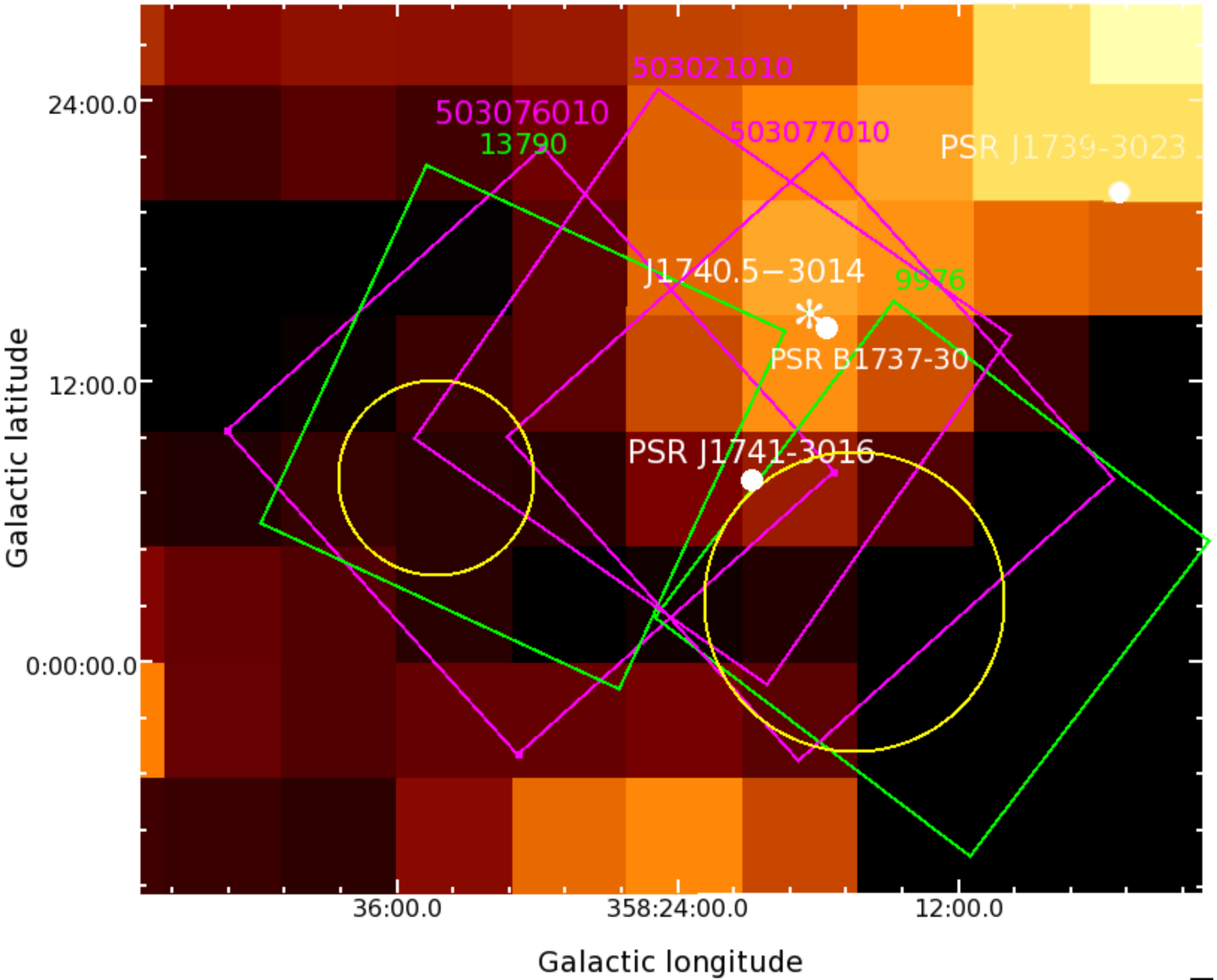}
\caption{Left: 
The \emph{Spitzer} {\citep{2006AJ....131.2859F}} image (8.0 $\mu$m) of the H.E.S.S. region. The green and magenta squares represent the \emph{CXO} and \emph{Suzaku} observations, respectively (with the corresponding ObsIDs on top). The yellow circles show the locations  and extension (see text for details) of the two bright regions within the H.E.S.S. source (smaller circle: J1741A, larger circle: J1741B). The three pulsars ( PSR B1737$-$30, PSR J1741$-$3016, and PSR J1739$-$3023) are shown with filled white circles. Right: The same region of sky as seen with {\sl INTEGRAL}  IBIS/ISGRI. In both images the position of Suzaku J1740 is shown as an asterisk (based on \citeauthor{2011PASJ...63S.865U} \citeyear{2011PASJ...63S.865U}).
}
\end{center}
\end{figure*}

There have been previous efforts to find a MW counterpart of HESS J1741--302. \cite{2010ecsa.conf..154M} and \cite{2011PASJ...63S.865U} have analyzed two {\sl Suzaku} observations, covering both J1741A and J1741B, to search for  X-ray counterparts.
 \cite{2010ecsa.conf..154M} claimed to have detected an X-ray counterpart  of  J1741A (see Figure 2 from \citeauthor{2010ecsa.conf..154M} \citeyear{2010ecsa.conf..154M}) with the XIS spectrum fitted by an absorbed power-law (PL) model with photon index $\Gamma=1.13\pm0.6$ and intervening hydrogen column N$_H=(3.95\pm2.7)\times10^{22}$\,cm$^{-2}$. The observed X-ray flux of {\sl Suzaku}  J1741.4--3006 (hereafter, Suzaku~J1741) in the 2$-$10 keV band,  $F_{\rm 2-10~keV}=3.2\times10^{-13}$\,erg\,s$^{-1}$\,cm$^{-2}$,  corresponds to a TeV to X-ray flux ratio of $F_{\rm 1-10~TeV}/F_{\rm 2-10~keV}\sim6$. Based on the high TeV to X-ray flux ratio, \citet{2010ecsa.conf..154M} suggested a hadronic origin of the TeV $\gamma$-ray emission from J1741A.  The extent of Suzaku J1741 ($\sim$ $5'$) is based on the X-ray contours shown in Figure 2 of \cite{2010ecsa.conf..154M}, which suggests that the source is resolved by {\sl Suzaku} XIS into diffuse emission.
 The {\sl Suzaku} observation covering J1741B also revealed an X-ray source, \emph{Suzaku} J1740.5$-$3014 (hereafter, Suzaku~J1740). The detected  periodicity of 432.1 $\pm$ 0.1 s together with the X-ray spectrum showing Fe I  K$\alpha$ emission, suggests that the source is a magnetic cataclysmic variable (CV), most likely an intermediate polar \citep{2011PASJ...63S.865U}.  Intermediate polars are not known (or expected) to produce VHE emission, making  Suzaku J1740 an  unlikely counterpart of the TeV source if the observed X-ray emission solely comes from the CV. 

Three pulsars located in the field of HESS J1741--302 are shown in Figure 1.  Of those three,  PSR B1737--30 (hereafter B1737), with a characteristic age of  $20.6 $ kyr, distance 5.5 kpc,  and spin-down energy loss rate  $\dot{E}=8.2\times 10^{34}$ erg s$^{-1}$ \citep{2010MNRAS.404..289Y, 2001MNRAS.322..715J}, is a powerful source of relativistic particles\footnote{This pulsar is also notable for the large number of  glitches ($\sim$20 in 20 years; see \citeauthor{2010MNRAS.404..289Y} \citeyear{2010MNRAS.404..289Y} and references therein).}.  \cite{2010ecsa.conf..154M} and \cite{2011PASJ...63S.865U} reported no X-ray emission coincident with the pulsar's location, however, the pulsar, if close enough to the CV position (which is only approximate\footnote{ The offset between the pulsar and the CV positions could be $\approx 0\farcm3$--$1.5'$ due to a large uncertainty in the {\sl Suzaku} pointing accuracy. The uncertainty is difficult to correct for due to the presence of only 3 very faint sources (in addition to {\sl Suzaku} J1740) in the XIS images and the very high density of 2MASS sources in this region. It is unclear from \citep{2011PASJ...63S.865U} how the $90''$ offset was determined from matching these X-ray sources with 2MASS stars despite the confusion caused by the high density of NIR sources.}), could be masked in the XIS images by the bright CV emission. 
Because of the large ($\approx12'$) offset of the pulsar from the peak of the TeV emission of J1741B,  one  needs to investigate and exclude other possible counterparts before declaring J1741B  a relic PWN candidate associated with B1737. Another pulsar, PSR J1741--3016, which is significantly closer to J1741B, has a characteristic age of  $3.3$ Myrs, distance 5.02 kpc, and  much smaller $\dot{E}=5.2\times10^{31}$ ergs s$^{-1}$ \citep{2002MNRAS.335..275M,1993ApJ...411..674T}. Therefore,   this pulsar it is too old to sustain a detectable PWN both in X-rays and TeV \citep{2009arXiv0907.0574T}. The PWN of the third pulsar, PSR J1739-3023, with a characteristic age of $159$ kyr , distance 3.41 kpc, and $\dot{E}=3.0 \times 10^{35}$ erg s$^{-1}$ \citep{2002MNRAS.335..275M, 1993ApJ...411..674T} could be the source of TeV $\gamma$-rays but the pulsar is very offset ($\sim 24'$ from J1741B; see Fig.~1) which makes it a very unlikely counterpart unless it is a very fast moving pulsar or the ISM density is strongly non-uniform creating the right environment for 
 the host SNR reverse shock to expand asymmetrically (the SNR is not seen in the VLA survey images\footnote{http://archive.nrao.edu/nvas/}, which could be due to the advanced age of the SNR).

To better understand the nature of  J1741B we have carried out a {\sl Chandra X-ray Observatory} ({\sl CXO}) observation of J1741B and also retrieved an archival {\sl CXO} observation of the J1741A field. Here we present the analysis of both observations including the MW analysis of the X-ray sources seen in the J1741A/B fields. We classify detected X-ray sources and investigate the origin of different sources seen in this region at lower energies. 
 We discuss whether any of these sources could be responsible for the TeV emission.

This paper is organized as follows. Section 2 summarizes the X-ray observations and data reduction, Section 3 discusses the search for MW counterparts of the detected X-ray sources,  Section 4 presents the results from the analysis of the MW data, and  Section 5 summarizes our findings. Section 6 is an appendix describing the details of our MW classification tool.

\section{CXO Observations and Data Reduction}

Two {\sl CXO} observations, ObsIDs 9976 (PI Tibolla; 19.7 ks exposure) and 13790 (PI Kargaltsev; 44.4 ks exposure), of HESS J1741--302 were used in our analysis covering the fields of J1741B and J1741A, respectively.
 In both observations the data were taken with the ACIS-I instrument operated in ``Very Faint" Timed exposure mode. We processed the data using the {\sl CXO} Interactive Analysis of Observations (CIAO\footnote{http://cxc.harvard.edu/ciao/index.html}) software (version 4.6) and {\sl CXO} Calibration Database (CALDB) version 4.5.9. We restricted our analysis to the energy range of  0.5--7 keV. We used CIAOs Mexican-hat wavelet source detection routine {\sl wavdetect} \citep{2002ApJS..138..185F} to detect X-ray sources and measure their coordinates in the {\sl CXO} images (see Table~1 and Figure~2). CIAO's task {\sl srcflux} was used to calculate the observed source fluxes. For the brightest sources, the X-ray spectra and responses were extracted using standard CIAO procedures.  The X-ray spectra were binned to a minimum of 10 counts per bin before fitting. Fits to X-ray spectra were performed using XSPEC 12.8.2.
 
  In order to look for extended sources the CIAO tool vtpdetect was run on the 0.5$-$7 keV band image. No extended sources were found. We also created a ``fluxed'' image by subtracting the point sources and replacing them with the local background using the dmfilth CIAO tool\footnote{http://cxc.harvard.edu/ciao/threads/diffuse\_emission/} and then dividing by the exposure map but no signs of diffuse emission are seen at various binning and smoothing scales.

We have detected 12 and 7 X-ray point sources in the fields of J1741A  and J1741B, respectively, at a  significance level $> 4\sigma$ (Table~1). For the brightest sources in each field  the spectra were extracted from a $3''$ radius circle for source 13 and 7$''$ radius circles for the other sources because they were imaged at large off-axis angles (see Table 1).  For each of these sources spectra we fitted absorbed PL and absorbed blackbody (BB) models (except for sources 9 and 10; see below) with N$_H$ fixed to the galactic value. The spectrum of source 9 was fitted with a MEKAL model \citep{1986A&AS...65..511M}, instead of a PL model,  with the absorption as a free parameter (because other fits failed). All fits used the XSPEC {\tt phabs} model \citep{1992ApJ...400..699B} for interstellar absorption. Best-fit parameters for each model can be found in Table 2, while the spectra and the fits are shown in Figure 3.

\begin{figure*}
\begin{center}
\includegraphics[scale=0.37]{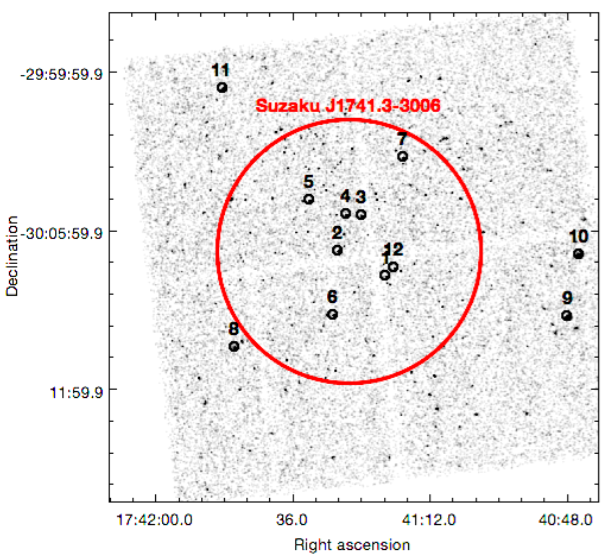}
\includegraphics[scale=0.38,trim= 0 1 0 0]{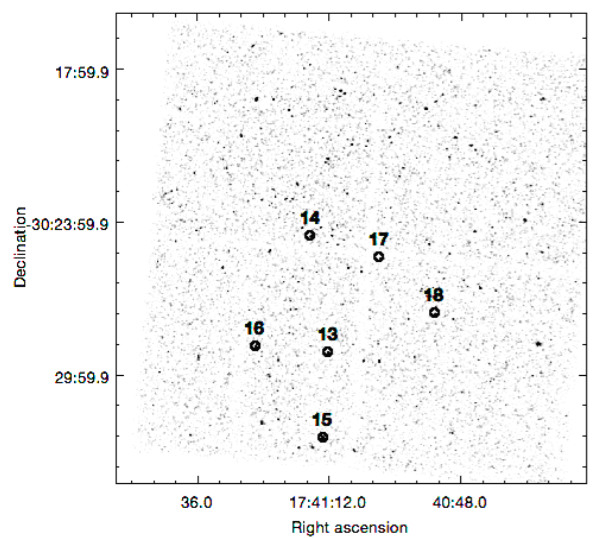}
\caption{\emph{CXO}/ACIS-I images of ObsId 13790 (J1741A field; left panel) and 9976 (J1741B field; right panel) with 19 X-ray sources detected (Table~1)  in the 0.5$-$7.0 keV energy band. North is up right is east.  Gaussian smoothing with a kernel radius of 3$''$ was applied to each image. The red circles show the extent of the source detected by {\sl Suzaku} \citep{2010ecsa.conf..154M}. 
}
\end{center}
\end{figure*}

\section{MW Counterpart Search}

We collected the MW properties (0.5-2.0 and 2.0-7.0  keV fluxes, two hardness ratios\footnote{Hardness ratios are calculated as HR2=($F_{1.2-2}$-$F_{0.2-1.2}$)/($F_{1.2-2}$+$F_{0.2-1.2}$) and HR4=($F_{2-7}$-$F_{0.5-2}$)/($F_{2-7}$+$F_{0.5-2}$), where $F_{x-y}$ are the observed fluxes in the respective energy band $x$$-$$y$ keV.  These energy bands were chosen to be compatible with those used in the Chandra Source Catalog \citep{2010ApJS..189...37E}.}, optical, NIR, and IR photometry) of the 19 sources detected in the ACIS-I images. The MW photometry was taken from optical (USNOB; \citeauthor{2003AJ....125..984M} \citeyear{2003AJ....125..984M}), NIR (Two Micron All Sky Survey; 2MASS; \citeauthor{2006AJ....131.1163S} \citeyear{2006AJ....131.1163S}), and IR ({\sl Spitzer}, {\sl WISE}; \citeauthor{2006AJ....131.2859F} \citeyear{2006AJ....131.2859F}; \citeauthor{2011wise.rept....1C} \citeyear{2011wise.rept....1C}) surveys. Only 8 of the 19 total sources in both fields were found to have MW counterparts in these surveys.  The optical, NIR, or IR source was considered to be a counterpart of an X-ray source if their positions are within $2''$ of each other. For each survey we calculate the chance superposition based on the average optical/NIR/IR source densities in the field ($\rho$ = 0.00249, 0.0108, and 0.00187 sources/arcsec$^{2}$ in the USNO-B1, 2MASS, and {\sl WISE} surveys, respectively). We calculate the probability of finding zero field sources in the circle of radius $r = 2''$, as $P = \exp(-\rho\pi r^2)$. Therefore, the probabilities of each counterpart to be due to chance coincidence is $1-P$ = 3.1\%, 12.7\%, and 2.3\% (for USNO-B1, 2MASS, and {\sl WISE} surveys, respectively). The chance coincidence probability is highest for 2MASS where we expect up to 2 of the cross-matches to possibly be spurious. None of the 19 sources have more than one MW counterpart within $2''$ of their X-ray position.  The IR/NIR/optical magnitudes of the potential counterparts are listed in Table~3.  

We used this MW information together with the measured X-ray properties to classify these 19 sources.  Using two different machine-learning methods described in the Appendix, 10 sources were classified with $>$70$\%$ confidence by at least one of the algorithms and, of these, 8 had consistent confident\footnote{ Note that the
calculated confidences  do not include the
uncertainties associated with the counterpart confusion,  X-ray flux determination, and hardness ratio calculation. See the Appendix for the confidence calculation details.} ($>$70$\%$) classifications by both algorithms. Below we only report the results consistent across both methods. The 8 classifications with confidences $>70\%$ shown in Table 1, include 4 AGN and 4 stars.

Figure 1 shows IR (\emph{Spitzer}; left panel) and hard X-ray (\emph{INTEGRAL}; right panel) images of the HESS J1741--302 region.  The ISGRI image in the 13--80 keV band was obtained using the INTEGRAL data processing pipeline in HEAVENS\footnote{ http://isdc.unige.ch/heavens/}\citep{2010int..workE.162W} based on a 7.2 Ms exposure. Near the position of  \emph{Suzaku} J1740 a clear enhancement is  seen in the  {\sl INTEGRAL}/ISGRI image. The 8~$\mu m$ IR image suggests that there is star forming activity in the J1741A field. There also appears to be a bright star surrounded by a bubble near the center of the J1741A source. It is known as Wray 17-96 recently identified as a candidate luminous blue variable (LBV) within a large spherical ejecta shell at the distance of $\sim4.5$ kpc \citep{2002ApJ...572..288E}. No X-ray counterpart is seen in the \emph{CXO} images nor is such a star expected to produce TeV emission. In the field of J1741B there is a known bright Wolf-Rayet (WR) binary (WR 98a), which has a relatively bright X-ray counterpart (see below). We also inspected the archival VLA images,  however, within both {\sl CXO} fields we only found a couple of faint radio point sources lacking  classifications and  IR or X-ray counterparts.
In addition to the low-energy surveys, we have searched the 3FGL catalog \citep{2015ApJS..218...23A}  and 1FHL catalog \citep{2013ApJS..209...34A}. The closest GeV sources are located at a distance of  $46'$ from the center of J1741A  (R.A.=17$^{h}$41$^{m}$40$^{s}$ Decl.=$-$30$^{\circ}$05$''$00$'$) and $31'$ from the center of J1741B (R.A.=17$^{h}$41$^{m}$17$^{s}$, Decl.=$-$30$^{\circ}$23$''$00$'$), implying that no GeV counterpart of HESS J1741-302 is detected.

\begin{table}
\caption{{\sl CXO} sources in fields of HESS~J1741-302 A and B (see Figures 2 and 3).}
\begin{center}
\begin{tabular}{ccccccccc}
\tableline
$\#$ & CXOU\tablenotemark{a} & Field & R.A.\tablenotemark{b} & Decl.\tablenotemark{b} & $F$\tablenotemark{c} & Counts\tablenotemark{d} & HR\tablenotemark{e}& Class\tablenotemark{f} (\%) \\
\tableline
 1 &  J174119.6-300745 & J1741A & 17$^{h}$41$^{m}$19$^{s}$.62 & -30$^{\circ}$07$''$45$'$.1 & 1.8$\pm$0.5 & 33 & 0.95 & AGN (84) \\
 2 &  J174128.0-300647 & J1741A & 17$^{h}$41$^{m}$28$^{s}$.00 & -30$^{\circ}$06$''$47$'$.7 & 4.4$\pm$0.9 & 71 & 0.99 & ?\tablenotemark{g} \\
 3 &  J174123.9-300528 & J1741A & 17$^{h}$41$^{m}$23$^{s}$.95 & -30$^{\circ}$05$''$28$'$.2 & 2.6$\pm$0.8 & 36 & 0.91 & AGN (87) \\
 4 &  J174126.6-300525& J1741A & 17$^{h}$41$^{m}$26$^{s}$.61  & -30$^{\circ}$05$''$25$'$.5 & 5$\pm$1 & 90 & 0.95 & AGN (78) \\ 
 5 &  J174133.0-300453& J1741A & 17$^{h}$41$^{m}$33$^{s}$.04 & -30$^{\circ}$04$''$53$'$.2 & 3.1$\pm$0.8 & 40 & 0.99 & ? \\
 6 &  J174128.8-300912& J1741A & 17$^{h}$41$^{m}$28$^{s}$.83 & -30$^{\circ}$09$''$12$'$.5 & 2.4$\pm$0.8 & 33 & 0.95 & ? \\
 7 &  J174116.6-300315 & J1741A & 17$^{h}$41$^{m}$16$^{s}$.66 & -30$^{\circ}$03$''$15$'$.1  & 1.7$\pm$0.5 & 43 & 0.82 & ? \\
 8 &  J174146.1-391926& J1741A & 17$^{h}$41$^{m}$46$^{s}$.16 & -30$^{\circ}$10$''$26$'$.3 & 3.3$\pm$0.9 & 47 & 0.99 & ? \\
 9 &  J174047.8-300916 & J1741A & 17$^{h}$40$^{m}$47$^{s}$.84 & -30$^{\circ}$09$''$16$'$.0 & 6.5$\pm$0.8 & 201 & -0.54 & STAR (95)\\ 
10 &  J174045.8-300654 & J1741A & 17$^{h}$40$^{m}$45$^{s}$.85 & -30$^{\circ}$06$''$54$'$.3 & 7$\pm$1 & 160 & 0.52 & ? \\
11 &  J174148.1-300039 & J1741A & 17$^{h}$41$^{m}$48$^{s}$.18 & -30$^{\circ}$00$''$39$'$.3 & 9$\pm$2 & 89 & 0.91 & ? \\
12 &  J174118.1-399725 & J1741A & 17$^{h}$41$^{m}$18$^{s}$.19 & -30$^{\circ}$07$''$25$'$.5 & 0.6$\pm$0.2 & 33 & -0.87 & STAR (92) \\
13 &  J174112.1-302908 & J1741B & 17$^{h}$41$^{m}$12$^{s}$.13 & -30$^{\circ}$29$''$08$'$.4 & 11$\pm$2 & 113 & 0.91 & ? \\
14 &  J174115.2-302434 & J1741B & 17$^{h}$41$^{m}$15$^{s}$.25 & -30$^{\circ}$24$''$34$'$.8 & 2.7$\pm$0.6 & 68 & -0.87 & STAR (97) \\
15 &  J174113.0-303230 & J1741B & 17$^{h}$41$^{m}$13$^{s}$.04 & -30$^{\circ}$32$''$30$'$.7 & 9$\pm$1 & 104 & 0.61 & ? \\
16 &  J174125.3-302853 & J1741B & 17$^{h}$41$^{m}$25$^{s}$.30 & -30$^{\circ}$28$''$53$'$.8 & 11$\pm$3 & 54 & 0.91 & ? \\
17 & J174102.7-302525 & J1741B & 17$^{h}$41$^{m}$02$^{s}$.77 & -30$^{\circ}$25$''$25$'$.9 & 3$\pm$1 & 23 & 0.91 & AGN (85) \\
18 &  J174052.7-302737 & J1741B & 17$^{h}$40$^{m}$52$^{s}$.73 & -30$^{\circ}$27$''$37$'$.7 & 4$\pm$1 & 32 & 0.95 & ? \\
19 &  J174054.4-301933 & J1741B & 17$^{h}$40$^{m}$54$^{s}$.43 & -30$^{\circ}$19$''$33$'$.4 & 1.8$\pm$0.5 & 45 & -0.83 & STAR (97) \\
\tableline
\end{tabular}
\end{center}
\tablenotetext{a}{ Source name according to the standard {\sl Chandra} source naming convention.}
\tablenotetext{b}{ Coordinates are listed for the J2000 epoch.}
\tablenotetext{c}{Observed X-ray fluxes in the $0.5-7$~keV range in units of $10^{-14}$ erg~s$^{-1}$~cm$^{-2}$.}
\tablenotetext{d}{Number of counts in the $0.5-7$~keV range.}
\tablenotetext{e}{Hardness ratio calculated as $(H-M-S)/(H+M+S)$, where $S$, $M$, and $H$ are the number of counts in the 0.5--2.0~keV, 1.2--2.0~keV, and 2.0--7.0~keV bands, respectively.}
\tablenotetext{f}{Classification and its confidence according to the automated classification algorithm (see Appendix).}
\tablenotetext{g}{``?" denotes cases with less confident ($<$ 70\% confidence) classifications.}
\end{table}

\section{Results and Discussion}
\subsection{HESS J1741$-$302A}

Our analysis of the \emph{CXO} image (ObsID 13790), which covers the J1741A region, reveals a number of X-ray point sources with no trace of diffuse emission (Figure~2, left panel)  down to a surface brightness limit of $\sim$2.5$\times$10$^{-15}$ erg cm$^{-2}$ s$^{-1}$ arcmin$^{-2}$ in the 0.5$-$7 keV band. This limit is based on an absorbed PL model (with $\Gamma=1.5$ and $N_H=1.47\times10^{22}$ cm$^{-2}$) that was used to simulate\footnote{with PIMMS v.4.7b; see http://cxc.harvard.edu/toolkit/pimms.jsp } a flux corresponding to the measured $1\sigma$ excess of background-subtracted counts from the Suzaku J1741 region (shown in Figure 2) after the point source removal (see Section 2).  The large apparent extent of the X-ray source in the \emph{Suzaku} XIS images can be explained by multiple faint point sources seen in the ACIS-I image (labelled 1, 2, 3, 4, 5, 6 and source 12 in Figure 2, left panel), which have been smeared out by the broad PSF of the \emph{Suzaku} XRT. The total observed flux from sources 1 to  6 and 12 is $F_X\approx2\times10^{-13}$\,erg\, cm$^{-2}$\, s$^{-1}$, comparable to the \emph{Suzaku} XIS  flux  reported for the {\sl Suzaku}  source \citep{2010ecsa.conf..154M}. Our MW classification suggests that sources 1, 3, and 4 are  AGN (classification confidence is 84\%, 87\%, and 78\%, respectively), which could, in principle, be responsible for the TeV $\gamma$-rays. However, the AGN\footnote{http://tevcat.uchicago.edu/} that emit at TeV energies typically have X-ray to TeV $\gamma$-ray flux ratios 2--3 orders of magnitude higher than those derived from the X-ray fluxes for these AGN candidates ($ F_{\rm 2-10 \ keV}/F_{\rm TeV} =$0.01--0.03), thus making these sources unlikely counterparts for the TeV source (see \citeauthor{2007ApJ...670..643K} \citeyear{2007ApJ...670..643K}).

Our classification algorithm was unable to confidently classify  source 2 (due to missing MW information in 6 bands) while the source is too faint (71 counts in ACIS-I) to meaningfully fit its X-ray spectrum. 
However, there is a clear deficit of soft X-rays from this source, suggesting that the spectrum is either very hard (typical of AGN and some X-ray binaries) or, alternatively, the source is intrinsically absorbed (could be either a remote quiescent XRB or an AGN). If this source is an AGN, it still has too low of an X-ray to TeV flux ratio (F$_{X-ray}$/F$_{TeV}$$=0.02$) to be a plausible candidate of the TeV source (see above). Alternatively, the source could be an XRB on the other side of the Galaxy, with large absorption making it appear faint and hard. However, typical $\gamma$-ray binaries have X-ray to TeV flux ratios between $\sim$ 0.5--12 \citep{2014AN....335..301K},  making this source an unlikely $\gamma$-ray binary candidate. Furthermore, a faint NIR counterpart is found in the 2MASS survey for this source (see Table 3) with no optical counterpart, suggesting that this could be an extincted and remote low mass X-ray binary (LMXB). Similar systems are known (e.g., XTE J1550-564; see Table 2 in \citeauthor{2011A&A...529A...3C} \citeyear{2011A&A...529A...3C}), but are typically much brighter in X-rays (although there are some exceptions; see e.g. \citealt{2014MNRAS.444..902A}). In any case, LMXBs are not known to emit TeV $\gamma$-ray emission.

We have also extracted the spectra of the two brightest sources ( 9 and 10, in Tables 1 and 2) in this field and fitted them. The spectrum for source 9 can be described by a {\tt mekal} model (although the fit is not perfect) with a temperature of  $\sim0.7$ keV (see Table 2), suggesting the X-ray emission can come from coronal activity in a star.  Source 10's spectrum is well fit by an absorbed PL model with photon index $\Gamma=2.7\pm0.3$. For both sources 9 and 10 a BB model provided an unacceptable fit. The lack of an optical or NIR counterpart to source 10 rules out a nearby coronally active low-mass star classification for this source but also does not allow for a definitive classification. The source could be a remote and extincted AGN. A strongly absorbed middle-aged pulsar with thermal emission dominating below 2 keV is also an option.  In the latter case it could be that the relic PWN of this pulsar is powering the TeV emission. However, no extended radio emission is seen in the archival VLA images or the X-ray images near this source. The classification algorithm confidently identified source 9 as a star (95\% confidence), while source 10 was not decisively classified.

\begin{figure*}
\begin{center}
\includegraphics[scale=0.26,trim=0 0 0 0,angle=270]{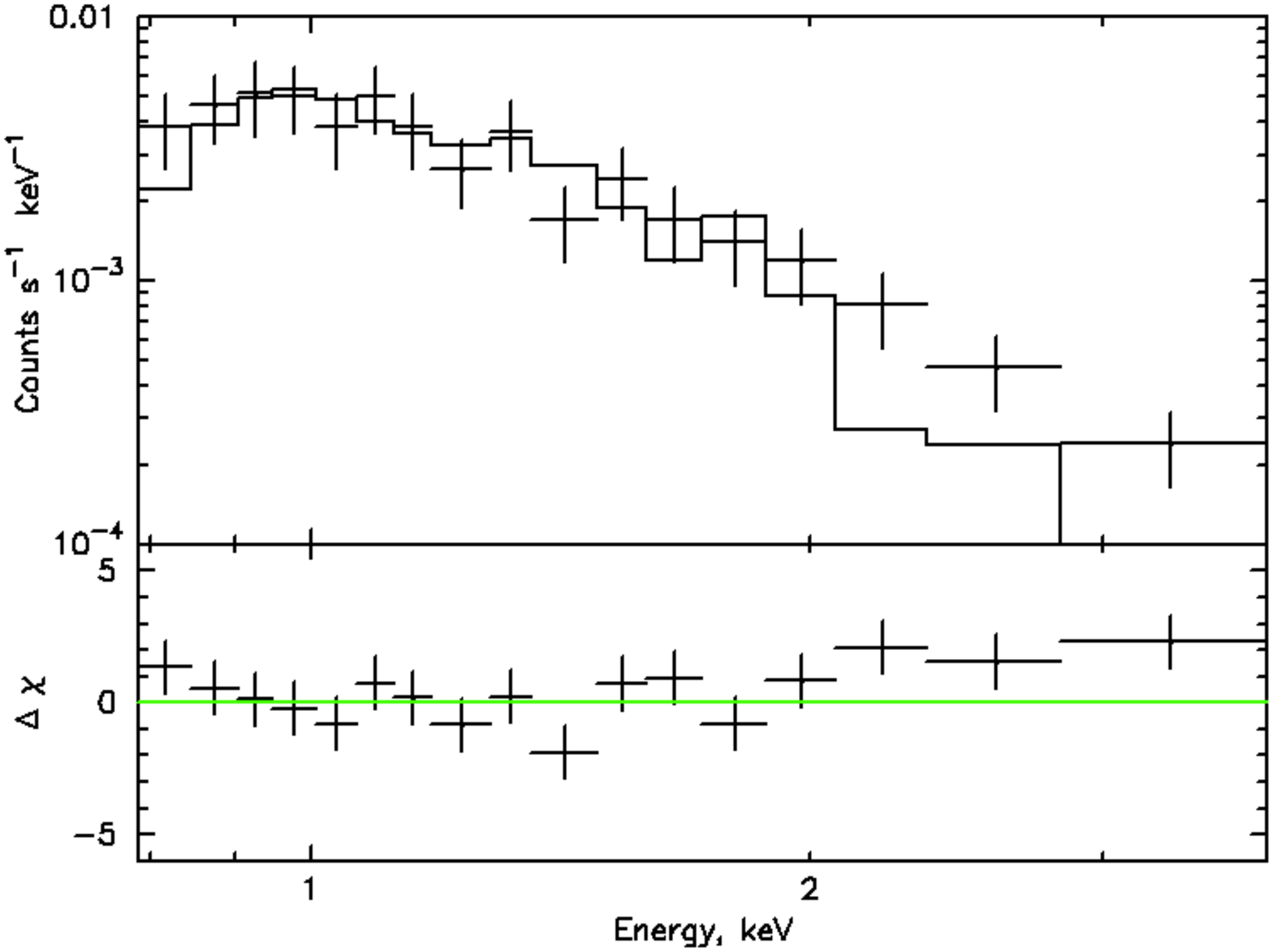}
\includegraphics[scale=0.26,trim=0 0 0 0,angle=270]{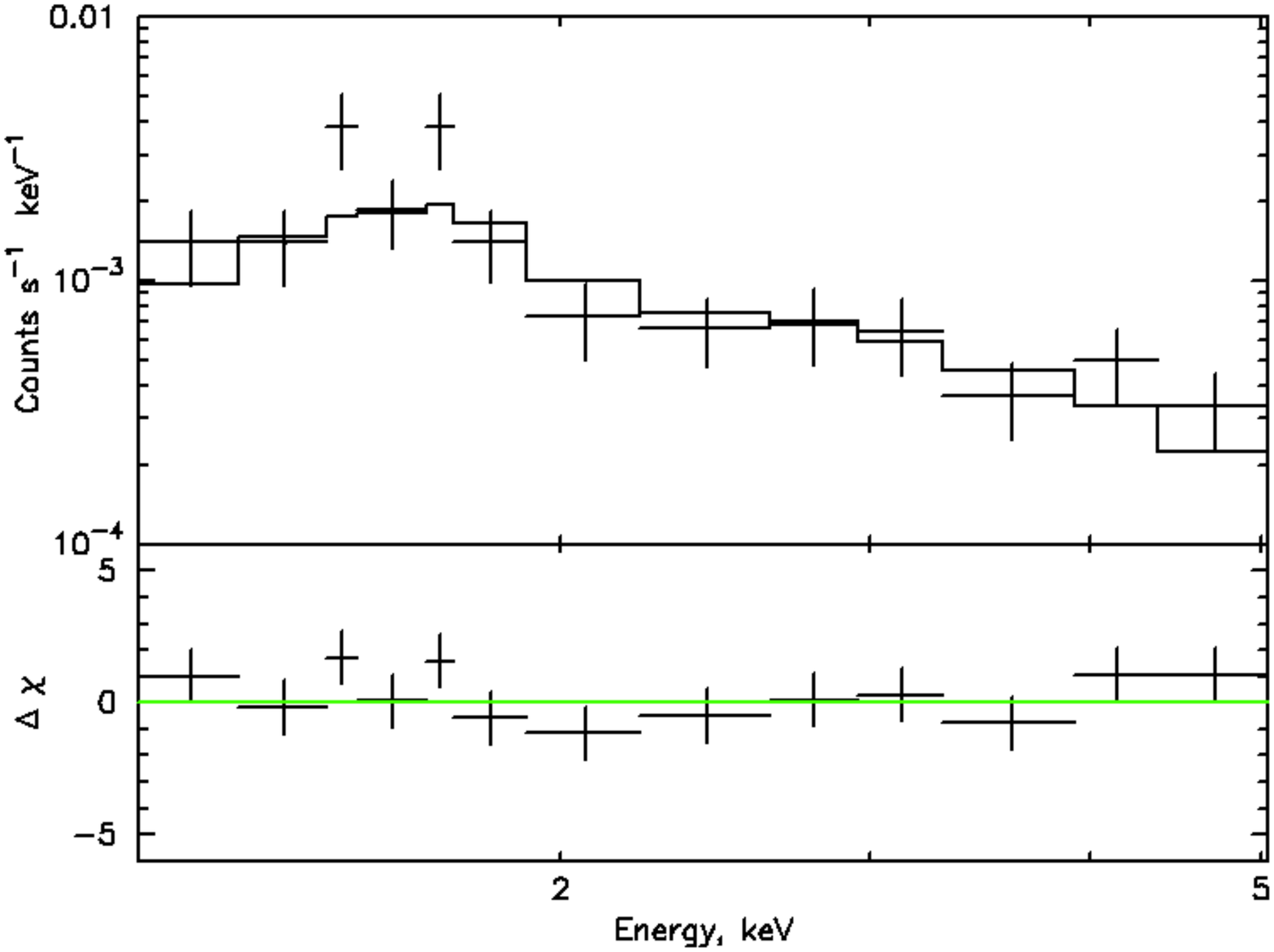}
\includegraphics[scale=0.26,trim=0 0 0 0,angle=270]{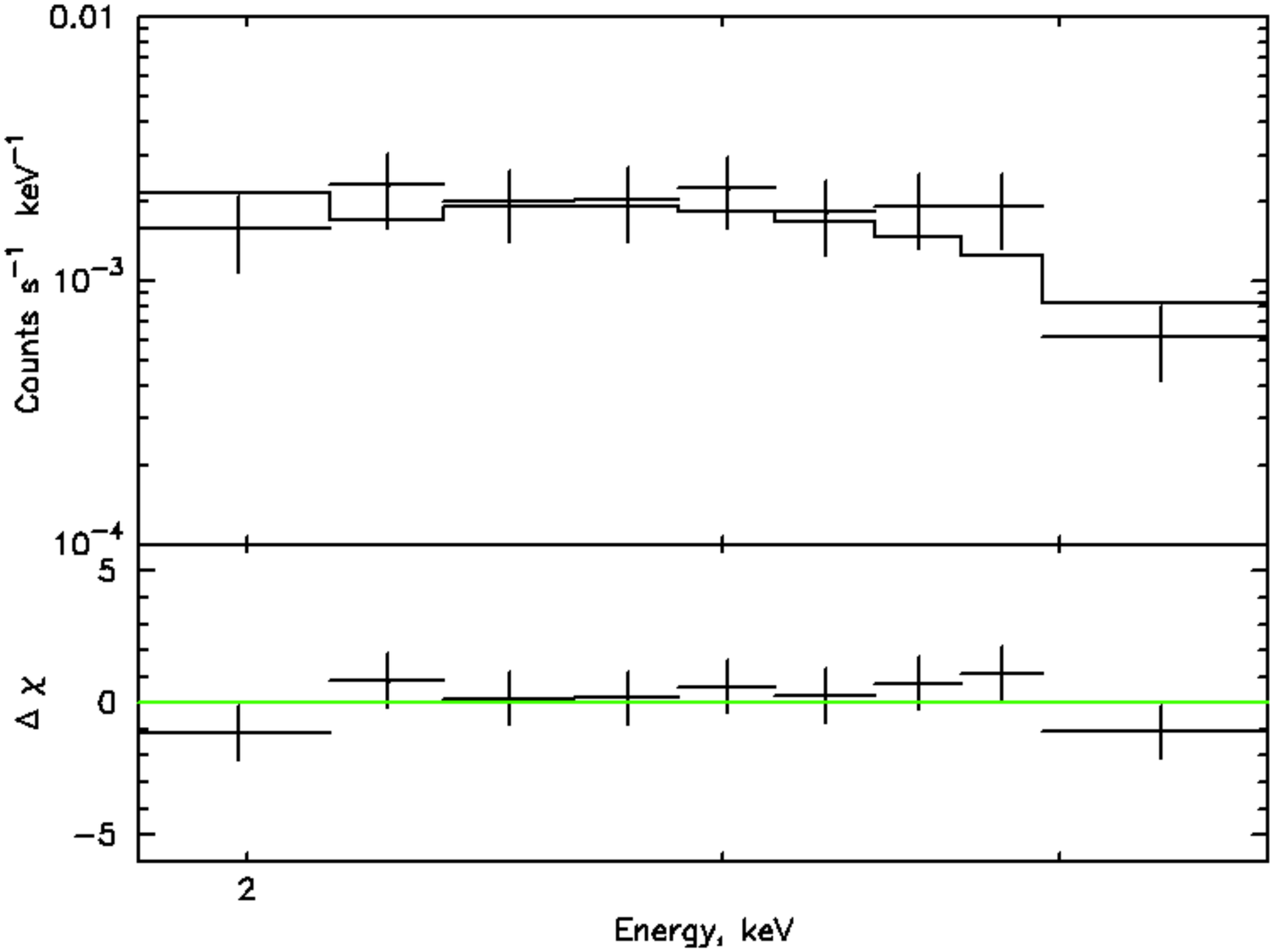}
\includegraphics[scale=0.26,trim=0 0 0 0,angle=270]{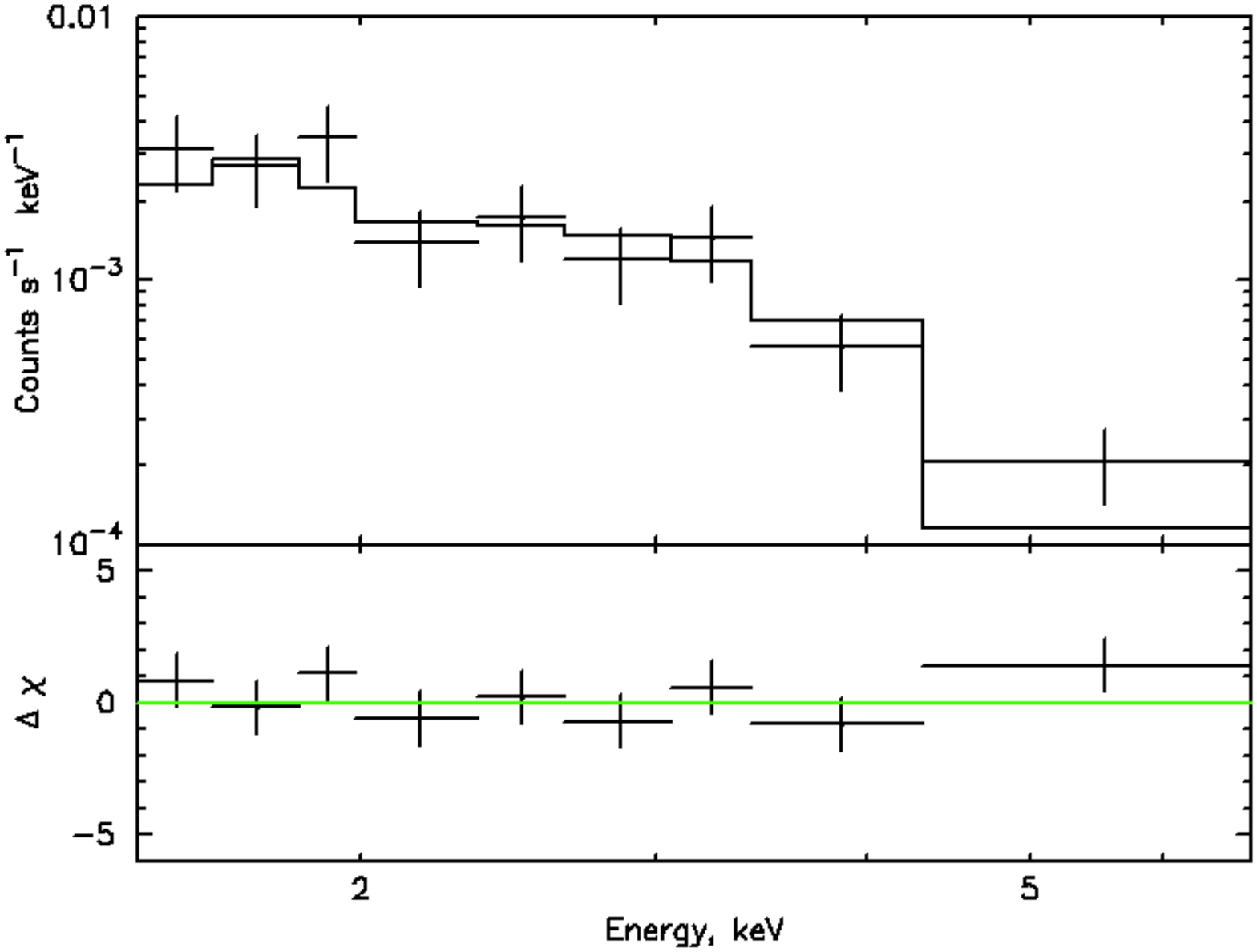}
\caption{ ACIS spectra for sources 9 (Mekal; upper left), 10 (PL; upper right), 13 (BB; lower left) and 15 (BB; lower right) in the fields of  J1741A/B (the ISM hydrogen column density is fixed at the Galactic value in this direction, $N_{H} =1.47\times 10^{22}$ cm$^{-2}$ \citep{1990ARA&A..28..215D}. The fit parameters can be found in Table 2. 
}
\end{center}
\end{figure*}

\begin{table}
\caption{Best-fit Model Parameters for Bright Source Spectra}
\begin{center}
\begin{tabular}{lccccc}
\tableline
Source & Model & N$_H$ & Norm\tablenotemark{a} & $kT$/$\Gamma$\tablenotemark{b}  & $\chi^{2}_{red}$ \\
\tableline
Source 9 & Mekal & (6.5$\pm$0.2)$\times$10$^{21}$ & (8$\pm$2$)\times 10^{-5}$ & 0.7$\pm$0.1 & 1.62  \\
Source 10 & PL & 1.47$\times10^{22}$\tablenotemark{\dagger} & (8$\pm$2$)\times 10^{-5}$ & 2.7$\pm$0.3 & 1.02  \\
Source 13 & BB & 1.47$\times10^{22}$\tablenotemark{\dagger} & (2.0$\pm$0.4$)\times 10^{-6}$  & 0.7$\pm$0.1 & 0.79  \\
Source 13 & PL & 1.47$\times10^{22}$\tablenotemark{\dagger} & (6$\pm$5$)\times 10^{-5}$  & 2.0$\pm$0.5 & 1.56  \\ 
Source 15 & BB & 1.47$\times10^{22}$\tablenotemark{\dagger} & (1.8$\pm$0.2$)\times 10^{-6}$  & 0.61$\pm$0.06 & 0.85  \\
Source 15 & PL & 1.47$\times10^{22}$\tablenotemark{\dagger} & (10$\pm$3$)\times 10^{-5}$  & 2.7$\pm$0.3 & 0.61 \\
\tableline
\end{tabular}
\end{center}
\tablecomments{The $1\sigma$ uncertainties are shown.}	
\tablenotetext{a}{The normalization for the BB model is R$^{2}_{km}$/D${^2}_{10}$ with R$_{km}$ defined as the source radius in km and D$_{10}$ as the distance to the source in units of 10 kpc. For PL, the normalization is defined in units of photons keV$^{-1}$ cm$^{-2}$ at 1 keV. The mekal model normalization is defined as 10$^{-14}$(4$\pi$[D$_A$(1+z)])$^{-1}$ $\int n_e n_H dV$ where D$_A$ is the angular diameter distance to the source in cm and n$_e$, n$_H$ are the electron and hydrogen densities respectively, in cm.}
\tablenotetext{b}{PL photon index or BB temperature in keV.}
\tablenotetext{\dagger}{Fixed at the galactic value \citep{1990ARA&A..28..215D}.}
\end{table}

\begin{table}
\caption{ Magnitudes of  potential MW counterparts to {\sl CXO} sources in fields of J1741A/B (see Figure 2).}
\begin{center}
\begin{tabular}{ccccccccrrr}
\tableline
\# & Field & b\tablenotemark{a} & r\tablenotemark{a} & i\tablenotemark{a} & j\tablenotemark{b} & h\tablenotemark{b} & k\tablenotemark{b} & w1\tablenotemark{c} & w2\tablenotemark{c} & w3\tablenotemark{c} \\
\tableline
 2 & J1741A & ... & ... & ... & 15.835 & 13.537 & 12.859 & ... & ... & ...   \\
 7 & J1741A & ... & 19.4  & 16.68 & 12.738 & 11.312 & 10.686 & ... & ... & ...  \\
 9 & J1741A & 15.14 & 13.54  & 12.6 & 11.985 & 11.446 & 11.115 & ... & ... & ... \\ 
12 & J1741A & 13.48 & 12.39 & 11.81 & 11.419 & 11.038 & 11.04 & ... & ... & ... \\
13 & J1741B & ... & 17.47 & 14.11 & 10.3 & 9.053 & 8.434 & 7.571 & 6.854 & 5.487\\
14 & J1741B & 16.04 & 14.34 & 13.67 & 12.887 & 11.796 & 11.787 & ... & ... & ... \\
15 & J1741B & 18.40 & 15.20 & ... & 9.143 & 6.505 & 4.332 & 6.552 & 6.188 & 2.824 \\
19 & J1741B & 12.64 & 12.06 & 11.86 & 10.944 & 10.75 & 10.631 & 9.567 & 9.585 & 7.422 \\
\tableline
\end{tabular}
\end{center}
\tablenotetext{a}{Magnitudes taken from the USNO-B optical catalog \citep{2003AJ....125..984M}.}
\tablenotetext{b}{Magnitudes taken from the Two-Micron All Sky Survey (2MASS; \citeauthor{2006AJ....131.1163S} \citeyear{2006AJ....131.1163S}).}
\tablenotetext{c}{Magnitudes taken from the Wide-Field Infrared Survey Explorer ({\sl WISE}; \citeauthor{2011wise.rept....1C} \citeyear{2011wise.rept....1C}).}
\tablecomments{``..." corresponds to the lack of counterpart at the corresponding wavelength within the $r=2''$ circle centered on the X-ray source.}
\end{table}

\subsection{HESS J1741$-$302B}

\citet{2010ecsa.conf..154M} report the discovery of X-ray source {\sl Suzaku} J1740, which is just outside the field of view in  both {\sl CXO} observations. The analysis of the X-ray properties  suggests that this source is likely a magnetic CV  \citep{2011PASJ...63S.865U}. On the sky, {\sl Suzaku} J1740 may be located in the immediate vicinity of PSR B1737$-$30. Therefore, PSR B1737$-$30 may be contributing to the observed X-ray emission from the binary, if it cannot be separated from the CV emission  due to the broad PSF  of {\sl Suzaku} XRT. The lack of {\sl CXO} data precludes pinpointing the precise position of the CV and confidently distinguishing the CV emission from that of PSR B1737$-$30 or its PWN. An additional {\sl CXO} observation is needed to isolate and study these two sources. 

The automated MW classification suggests that source 17 is an  AGN 
with fairly low  X-ray  flux, which makes it an unlikely counterpart to the TeV source (see above). The other two confidently classified sources (\# 14 and 19) appear to be stars. Four X-ray sources lack confident classifications. Of these, sources 13 and 15 are sufficiently bright to perform spectral fits (see Figure 3). Both sources are best-fit by an absorbed BB model and the best-fit parameters are given in Table 2. Neither source is confidently classified by our automated pipeline. The optical/NIR/IR magnitudes  for the possible MW counterparts of  sources 13 and 15   are listed in Table 3. 

 The candidate optical/NIR/IR counterpart of source 13 exhibits very red  colors (see Table 3). This dereddened r-band magnitude makes the optical to X-ray flux ratio unusually  large for  a typical AGN (see also Figure 4 in \citeauthor{2002Msngr.108...11H} \citeyear{2002Msngr.108...11H}).   Prior to any dereddening the optical/NIR/IR spectrum resembles that of a very cool star (L dwarf) which must then be very nearby (few tens of parsecs). However, the source does not exhibit any measurable proper motion\footnote{according to the PPMXL catalog of positions and proper motions by \cite{2010AJ....139.2440R}.} and the X-ray fit suggests substantial extinction (which perhaps could be intrinsic for a cool brown dwarf).  The uncertainty in the extinction, the unknown distance, and poor quality of the X-ray spectrum also leave a heavily obscured AGN as an option.  However, an r-band magnitude of 17.47, when dereddened by $E(B-V)=2.94$ (corresponding to  the total galactic absorption of N$_H$=1.47$\times 10^{22}$ cm$^{-2}$; \citep{1990ARA&A..28..215D}), becomes 10.3. This would correspond to an extremely large optical to X-ray flux ratio ($f_{\rm r}/f_{X}\sim700$) which is typical for a cool star and atypical for AGN. Therefore, the 
MW properties of source 13  are puzzling and it is not surprising that the automated classifier is confused. Of course, there always remains a small chance that the optical/NIR/IR source is not the true counterpart of the X-ray source 13.

Source 15 has a  possible optical and NIR counterpart and can be seen in the {\sl Spitzer} image in Figure 1.  It has a known classification of Wolf-Rayet star (known as WR 98a) \citep{1991ApJ...378..302C}. WR 98a is at a distance of $\sim 1.9$ kpc (\citealt{1999ApJ...525L..97M}) which corresponds to an X-ray luminosity of 3.8 $\times$ 10$^{31}$ ergs s$^{-1}$. Our classification algorithm was unable to confidently classify this source. However, WR 98a is surrounded by a dusty pinwheel nebula \citep{1999ApJ...525L..97M}. This is indicative of a tight binary which may contain two high-mass stars with strong colliding winds \citep{1999ApJ...525L..97M}. It is  likely that the automated  algorithm is confused by this, as it does not have colliding wind WR binaries as a separate class. Interestingly, it was suggested that such  binaries could be TeV sources (see \citealt{2008ApJ...685L..71A,2009arXiv0907.0399T,2010MmSAI..81..514P} and references therein) but none have been detected so far.

\section{Summary and Outlook}

Our MW classification of  19 sources from two  \emph{CXO}  observations did not yield an obvious candidate for the TeV emission from J1741A/B. Our automated machine learning classification algorithm relatively confidently classified 8 sources (4 AGN and 4 Stars). However,  HESS J1741-302 is not likely an AGN due to the  low X-ray to TeV flux ratios, while non-degenerate coronally active stars do not emit TeV $\gamma$-rays. Sources 2, 10 and 13  could be  obscured AGN or  remote XRBs while source 10 could also be a new middle-aged pulsar.
 Source 15 in the field of J1741B  has a known WR star (WR 98a)  within the {\sl CXO} error ellipse, which is likely a binary, raising the intriguing  possibility of the first detection of such a system in TeV $\gamma$-rays.  For the J1741A region we also cannot rule out TeV emission from the star forming clouds (see the Spitzer image in Figure 1) that provide a dense environment and enhanced IR photon background. This would still require a source of relativistic protons or electrons at the same distance. Suzaku J1741,  which appears to be extended in the {\sl Suzaku} image, and was previously mentioned as a plausible counterpart to J1741A, is resolved by {\sl CXO} into a number of faint point sources none of which looks particularly promising as a counterpart for the TeV emission. No diffuse emission is seen in the {\sl CXO} image.   The analysis of MW data also did not produce any credible counterparts. The {\sl INTEGRAL} flux enhancement near the position of Suzaku J1740 is consistent with the proposed magnetic CV nature, which implies no TeV emission. However, the PWN of the young radio pulsar B1737--30, located outside of the {\sl CXO} field of view, also remains a  possible counterpart for the TeV source. 
 
  Further observations of this field, including the three pulsars, are needed to look for possible offset PWNe from PSR J1739--3023 or PSR B1737--30. These observations would also enable us to spatially resolve the X-ray emission from PSR B1737--30 and cataclysmic variable Suzaku J1740. Future TeV observations will also help build enough statistics to determine whether there are two separate sources or only one larger extended source and further constrain the TeV spectrum. Until the contributions from the relic PWNe associated with the two pulsars are ruled out the source can hardly be on the list of ``dark accelerators", assuming that the "dark accelerator" term is reserved for TeV sources lacking any plausible counterparts in a reasonable vicinity.
 
\acknowledgments 

Support for this work was provided by the National Aeronautics and Space Administration through Chandra Awards GO2-13091X and AR3-14017X issued by the Chandra X-ray Observatory Center, which is operated by the Smithsonian Astrophysical Observatory for and on behalf of the National Aeronautics Space Administration under contract NAS8-03060. ES acknowledges support from The Science Academy (Bilim Akademisi, Turkey) under the BAGEP program.  We would like to thank the anonymous referee for carefully reading the paper and providing useful comments.

\section{Appendix} 
  Our supervised machine-learning pipeline relies on two supervised decision tree learning  algorithms and a training dataset. We use the See5 implementation\footnote{http://www.rulequest.com/see5-unix.html} of the C5  decision tree  algorithm (\citeauthor{1993cpml.book.....Q} \citeyear{1993cpml.book.....Q}) and a Random Forest\footnote{http://scikit-learn.org/stable/modules/ensemble.html} classifier \citep{RF}.

The algorithms determine the degree of similarity between the sources from the training dataset and unclassified sources using a number of MW parameters. All unclassified X-ray sources are cross-matched with MW catalogs in order to extract the MW parameters similar to those used in the training dataset described below (see Section 3 for the spatial selection criteria for MW counterparts). The following MW parameters are extracted: X-ray fluxes in 4 bands and two hardness ratios from the {\sl CXO} observations\footnote{The energy bands were chosen to be the same as those used in the Chandra Source Catalog.}, optical u,g,r,i,z magnitudes from the UNSO-B catalog \citep{2003AJ....125..984M}, NIR j,h,k magnitudes from the Two Micron All-Sky Survey (2MASS; \citealt{2006AJ....131.1163S}) and IR W1, W2, W3 magnitudes from {\sl Wide-field Infrared Survey Explorer} ({\sl WISE}; \citealt{2011wise.rept....1C}).

The training dataset we used here has $\sim$ 8,500 sources with 9 predefined object classes: {\sl (1) main sequence stars} (General Catalog of Variable Stars; \citealt{2009yCat....102025S}), {\sl (2) young stellar objects} (Chandra Orion Ultradeep Point Source Catalog and PAN-Carina; \citealt{2005ApJS..160..319G}, \citealt{2011ApJS..194...14P}), {\sl (3) AGNs} (Veron Catalog of Quasars \& AGN; \citealt{2010A&A...518A..10V}), {\sl (4) LMXBs} (Low-Mass X-Ray Binary Catalog, 2007; \citealt{2007A&A...469..807L}), {\sl (5) HMXBs} (Catalog of High-Mass X-Ray Binaries in the Galaxy; \citealt{2006A&A...455.1165L}), {\sl (6) cataclysmic variables} (CVs; Cataclysmic Variables Catalog, 2006, \citealt{2001PASP..113..764D}), {\sl (7) isolated neutron stars} (NSs; ATNF Pulsar Catalog; \citealt{2005AJ....129.1993M}), {\sl (8) binary non-accreting NS} (ATNF Pulsar Catalog), and {\sl (9) Wolf-Rayet stars} (The VIIth Catalog of Galactic Wolf-Rayet Stars; \citealt{2001NewAR..45..135V}). All of the AGN in our training dataset are located off of the galactic plane. Therefore, these AGN will look different from those in our observations due to galactic absorption. In order to correct for this, all AGN parameters in the training dataset were reddened using the total galactic absorption column in the direction of our observations (N$_H=$1.47$\times$10$^{22}$ cm$^{-2}$;  \citealt{1990ARA&A..28..215D}).

 A Laplace prescription for the estimation of the classification confidences\footnote{Note that the calculated confidences do not include the uncertainties in the X-ray flux, which can be significant for faint X-ray sources and will impact the HRs. Source confusion (i.e., the possibility of assigning a false MW counterpart to an X-ray source) is also not accounted for in the confidence calculation.}, $P$, has been adopted \citep{Chawla2003}. For each leaf node, $P=(TP+1)/(TP+FP+C)$, where $TP$, $FP$, and $C$ are true positives, false positives, and number of classes, respectively, for the leaf in which the source in question has landed based on its parameters.

To check the accuracy of both algorithms, we have used two fold cross-validation. This involves dividing the training dataset into two parts and then training the algorithms on this first half. The second half is then classified by the algorithms and then compared to their real classifications. The accuracy scores of C5 and Random Forest were $\sim$ 90\% and 93\%, respectively.


\end{document}